\documentstyle{article}
\include{epsf}
\begin{document}

\title{\bf Magnon-polaron and Spin-polaron Signatures in the Specific Heat and
Electrical Resistivity of $La_{0.6}Y_{0.1}Ca_{0.3}MnO_3$ in Zero Magnetic Field,
and the Effect of $Mn-O-Mn$ Bond Environment}

\author{M. Ausloos$^{1}$, L. Hubert $^{1}$, S. Dorbolo$^{1,2}$, A. Gilabert$^{3}$\\
and R. Cloots$^{4}$ \\
$^{1}$  SUPRAS, Institute of Physics, B5, \\ University of Li$\grave e$ge,
B-4000 Li$\grave e$ge, Euroland\\  $^{2}$  SUPRAS, Montefiore Electricity
Institute, B28,\\ University of Li$\grave e$ge, B-4000 Li$\grave e$ge, Belgium\\
$^{3}$ Laboratoire de Physique de la Mati$\grave e$re Condens\'ee, \\Universit\'e
de Nice-Sophia Antipolis, \\ Parc Valrose F-09016 Nice, Cedex 02, France \\$^{4}$
SUPRAS,  Institute of Chemistry, B6, \\ University of Li$\grave e$ge,  B-4000
Li$\grave e$ge, Belgium\\ }

\maketitle

\begin{abstract} $La_{0.6}Y_{0.1}Ca_{0.3}MnO_{3}$, an $ABO_{3}$ perovskite
manganite oxide, exhibits a non trivial behavior in the vicinity of the sharp
peak found in the resistivity $\rho$ as a function of temperature $T$ in zero
magnetic field. The various features seen on $d\rho/dT$ are discussed in terms of
competing phase transitions. They are related to the $Mn-O-Mn$ bond environment
depending on the content of the $A$ crystallographic site. A Ginzburg-Landau type
theory is presented for incorporating concurrent phase transitions. The specific
heat $C$ of such a compound  is also examined from 50 till 200 K. A log-log
analysis indicates different regimes. In the low temperature conducting
ferromagnetic phase, a collective magnon signature ($C \simeq T^{3/2}$) is found
as for what are called magnon-polaron excitations. A $C \simeq T^{2/3}$ law is
found at high temperature and discussed in terms of the fractal dimension of the
conducting network of the weakly conducting (so-called  insulating)  phase and
Orbach estimate of the excitation spectral behaviors. The need of considering
both independent spin scattering and collective spin scattering is thus
emphasized. The report indicates a remarkable agreement for the Fisher-Langer
formula, i.e. $C$ $\sim$ $d\rho/dT$ at second order phase transitions. Within the
Attfield model, we find an inverse square root relationship between the critical
temperature(s) and the total local $Mn-O-Mn$ strain.

\end{abstract}


\section{Introduction}

The properties of manganite's family $R_{1-x}A_{x}MnO_3$ compounds (where
$R=La,Y,Nd,Pr$ and $A=Ca,Sr,Ba,Pb$) with a $Mn^{3+}/Mn^{4+}$ mixed valence keep
attracting much attention of both experimentalists and
theorists.~\cite{review,7a,r0,1,2,3,4,5,6,7,8a,8b,9,10,11,12,13,14,us1,us2,usPRB,Huhtinen,bene,Castro,Okuda,Gordon,Cornelius}
In the doping range $0.2<x<0.5$, these compounds are known to undergo a double
magnetic and conductive phase transition under cooling from a paramagnetic (PM)
weakly conductive, usually called $insulating$ (I), state to a ferromagnetic (FM)
metallic-like (M) state. Thus a Curie temperature $T_C$ and a charge carrier
localization temperature $T_{MI}$ are respectively defined. The observable
difference between the two critical temperatures is usually attributed to the
quality of the sample~\cite{5,6,7,8a}. Nevertheless it could be ascribed to
intrinsic interaction interplays,  enhanced by the inhomogeneity content. No need
to recall that some anomalous expansion also takes place at the transitions,
\cite{Gordon} thus indicating a strong coupling between the lattice, spin, and
electronic degrees of freedom.

The magnetic localization of spin polarized carriers, forming so-called {\it spin
polarons} results in a diffusivity dominated charge carrier transport mechanism
below $T_C$ with a steadily increasing resistivity $\rho$ with increasing
temperature. However above $T_C$, the resistivity decreases and follows a
thermally activated Mott-like variable-range hopping law $\rho \propto
\exp(T_0/T)^{z}$ with $1/4\leq z \leq 1$. Despite a variety of theoretical
scenarios attempting to describe this phenomenon, practically all of them adopt
as a starting point the so-called double-exchange (DE) mechanism, which considers
oxygen mediated electron exchange between neighboring $Mn^{3+}/Mn^{4+}$ sites and
strong on-site Hund's coupling. In other words, the  mobility of the conduction
electrons between heterovalent $Mn/Mn$ pairs is supposed to be greatly enhanced
when the magnetic moments on adjacent $Mn$ ions are aligned. The mixed valency
also leads to the formation of small polarons, arising from $Mn/Mn$ valence
changes and Jahn-Teller (JT) distortions involving $Mn$ that leads to incoherent
hopping and high resistivity in the insulating phase. The estimated exchange
energy~\cite{11} $JS=45meV$ (where $S=2$ is an effective spin on a $Mn$ site),
being much less than the Fermi energy $E_F$ in these materials (typically,
$E_F=0.15eV$), favors an FM ground state.

The localization scenario,~\cite{13} in which $Mn$ oxides are modelled as systems
with both DE off-diagonal spin disorder and nonmagnetic diagonal disorder,
predicts a divergence of the electronic localization length $\xi (M)$ at some so
called M-I phase transition at $T_{MI}$. A critical spontaneous magnetization $M$
strength separates both phases, i.e. for $M$ small, $0<M<M_0$, the system is in a
highly resistive (insulator-like) state, while at low T, for $M>M_0>0$, the
system is in a low resistive, metallic-like, ferromagnetic phase. Within this
scenario, the Curie point $T_C$ is defined through the spontaneous magnetization
$M$ as $M(T_C)=0$, while the M-I transition temperature $T_{MI}$ is such that
$M(T_{MI})=M_0$ with $M_0$ being a fraction of the saturated magnetization $M_s$,
thereby with $T_{MI}$  $<T_{C}$.

The influence of magnetic fluctuations on electron-spin scattering near $T_{MI}$
and $T_{C}$ is expected to be rather important. They might easily tip a subtle
balance between magnetic and electronic processes in favor of either charge
localization or delocalization, and shift the relative position of $T_{MI}$  and
$T_{C}$. Moreover as in any phase transitions, a relevant question pertains to
the observation or not of critical fluctuations \cite{Huhtinen}, and the order of
the transition. \cite{Gordon} It seems that the ferromagnetic ordering is a
thermodynamic first-order transition, intrinsically broadened by a distribution
in $T_C$ \cite{Gordon}. We will give an interpretation of that broadening, and
its fine structure, from a chemical point of view.

An applied magnetic field $H$ enhances the FM order, thus reduces the spin
scattering and produces a negative  so-called giant magnetoresistivity (GMR). A
sharp peak occurs around $T_{MI}$. The localized spin disorder scattering is
thought to be highly responsible for the observed features in the GMR ~\cite{13}.
We will not report magnetic field effects here below.

In view of its charge carrier density sensitive nature, specific heat
measurements complement the traditional $\rho$ and GMR data and be used as a tool
for probing the intrinsic delocalization of the charge carriers below $T_{MI}$,
whence below $T_C$. The observed~\cite{14} giant magnetic {\it entropy change} in
manganites (produced by the abrupt reduction of the magnetization and congruent
to an anomalous thermal expansion near the Curie point) gives another reason to
utilize the specific heat data in order to get an additional information on the
underlying interaction mechanisms in these materials as well as on excitations
present in the vicinity of the critical temperature(s)
\cite{Castro,Okuda,Gordon,Cornelius}.

On the other hand, substitution on the $A$ (or $R$) site is known to modify the
phase diagram through cation size effects leading toward either a charge-ordered
(CO) or an antiferromagnetic (AFM) instability ~\cite{6}. In particular, $Y$
substitution is responsible for weakening the system's robustness against strong
AFM fluctuations, developed locally within the ordered FM matrix thereby shifting
$T_C$. Some difference in $T_C$ and $T_{MI}$ positions related to modified
collective excitations can be expected.

In $La_{0.6}Y_{0.1}Ca_{0.3}MnO_3$, \cite{r0} the negative GMR $\Delta \rho$
observed  at $B=1T$ shows a fine symmetry around $T_{0}=160K$. This suggests a
usual transport mechanism controlled by fluctuations, in a polycrystalline or
inhomogeneous system, since it is known that the transition should be very sharp
for single grains  \cite{bene}. There are indications~\cite{10} that the
observable GMR $\Delta \rho (T,B)$ scales with the magnetization $M$ in the
ferromagnetic state and follows an $M^{2}$ dependence in the paramagnetic region
implying thus some kind of universality in the magneto-electrical transport
properties below and above $T_C$, as in metals
\cite{Mott,VPT,DGF,Kasuya,MAKD,MAbk}. Strong magnetic (and charge) fluctuations
are thought to be triggered by $Y$ substitution and further enhanced by the
magnetic field. The \cite{us1,us2} data  was interpreted in terms of nonthermal
spin hopping and magnetization $M$ dependent charge carrier localization leading
to $\Delta \rho =-\rho _s\left( 1-e^{-\gamma M^{2}}\right )$ with
$M(T,B)=CB/|T-T_C|^{\nu}$ \cite{us1}. This formula generalizes the usual law for
independent spin scattering in metals \cite{Mott,VPT,DGF}.

In the present paper we report and discuss some analysis of typical results on
the specific heat for a $La_{0.6}Y_{0.1}Ca_{0.3}MnO_3$ sample from the same batch
previously used to measure the GMR and magneto-thermoelectric power (MTEP)
~\cite{us1,us2}. A wide temperature interval ranging from $20K$ to $300K$ has
been investigated with great care, using a PPMS from Quantum Design
\cite{stephaneprb} when it works. The method is a semi-adiabatic experiment.  We
observe contributions from the collective spin reservoir.  For the electrical
resistivity the same care was taken as in investigations aimed at measuring
critical exponents \cite{sousa,laurent}.

From the theoretical point of view we adopt the main ideas of the microscopic
localization theory~\cite{13} and can construct a phenomenological free energy
functional of Ginzburg-Landau (GL) type which describes the temperature behavior
of the spontaneous magnetization in the presence of strong localization effects.
Calculating the background and fluctuation contributions to the total
magnetization within the GL theory, the localization related magnetic free energy
leads to the specific heat through

\begin{equation} C = - \frac {\partial ^{2} \cal{F}} {\partial T^{2}}.
\end{equation}

It is also known \cite{FisherLanger} that when (critical) fluctuations are
important, the specific heat $C$ and $d\rho/dT$ contain the same temperature
dependent kernel, thus

\begin{equation} C \simeq \frac {\partial  \rho} {\partial T}. \end{equation}

\section{Experimental results}

The $La_{0.6}Y_{0.1}Ca_{0.3}MnO_3$ samples were prepared from stoichiometric
amounts of $La_{2}O_3$, $Y_{2}O_3$, $CaCO_3$, and $MnO_2$ powders, among many
other cases. The mixture was heated in air at $800$$^{\circ}$C  for 12 hours to
achieve decarbonation and  was pressed at room temperature under $10^3kG/cm^2$ in
order to obtain parallelipedic pellets. A slow (during 2 days) annealing and
sintering process was made from $1350$$^{\circ}$C to $800$$^{\circ}$C in order to
preserve the  stoichiometry, though no perfect sample homogeneity is claimed.

A small bar (length $l=10mm$, cross section ${\cal S}=4mm^2$) was cut from one
pellet. The electrical resistivity $\rho (T)$ was measured using the conventional
four-probe method, taking a data point every 0.5K. To avoid Joule and Peltier
effects, a DC current $I=1mA$ was injected (as a one second pulse) successively
on both sides of the sample. The voltage drop $V$ across the sample was measured
with high accuracy by a $KT182$ nanovoltmeter. Measurements {\it in a magnetic
field} indicated a magnetoresistance (MR) $\Delta \rho (T,H)=\rho (T,H)-\rho
(T,0)$ as shown in Fig. 1 in ref.\cite{us1}. The negative MR $\Delta \rho (T,H)$
shows a peak  at some temperature $ca.$ $T^{*}=170$ K. The magneto thermopower
(MTEP) $S$ was also measured as reported elsewhere ~\cite{us2}.

However further examination of $d\rho/dT$ shows some fine structure, as seen in
Fig.1. The inflexion point of $\rho$ is precisely defined through $d\rho/dT$ at $
T= 149$ K, the maximum in $\rho$ occurs for $d\rho/dT=0$ at $T=170$ K. A second
inflexion point occurs at 188 K. Moreover one observes a singularity in
$d\rho/dT$ above the inflexion point, i.e. at $T= 158$ K, and $\lambda$-like
peaks at 94, 110, and 130 K. This may be reminding us of $d\rho/dT$ behavior at
ferromagnetic transitions in metals and suggests the existence of specific
energies.

We were able in \cite{us1} to successfully fit the $\Delta \rho (T,B)$ data,
where  $\Delta$ means the deviation from the B=0 case, for the whole temperature
interval with \begin{equation} \Delta \rho (T,B)=-A[1-e^{-\beta (T)}],
\end{equation} where \begin{equation} \beta (T)=\beta _0\left[
\frac{T_0}{T-T_0}\right ]^{2\nu}, \end{equation} in which $A$, $\beta _0$ and
$\nu$ are temperature-independent parameters, under the condition $T_{0} =
T^{*}$. Furthermore we observe that the $d\rho/dT$ behavior rather looks like the
change in resistivity occurring near most anti- and/or ferromagnetic
order-disorder phase transitions \cite{DGF,MAbk,FisherLanger,geldart1,geldart2}.

Next, Fig. 2 shows a typical unsmoothened run for the temperature behavior of the
specific heat $C(T)$, from which the ratio $C(T)/T$ is obtained (Fig. 1).
Practically, a 10 mg sample has been placed on a sample holder composed of a
little paddle (3 $\times$ 3 mm$^2$) in teflon.  A heather placed on the back of
the paddle increases the temperature of the sample during 3 s; a data point is
taken every 1.5K. Both thermal answers of sample holder and sample are recorded
and fitted by exponential laws for which the characteristic time is related to
the thermal capacity of the system. The addendum contribution has to be first
measured and substracted from the raw data. We stress that this is done for the
same sample as that used for measuring the MR. It is observed that singularities
occur at the $same$ temperatures for $C$ and $d\rho/dT$. In the intermediate
temperature regime hereby examined (from 108 till 156K) the rather flat though
quite bumpy behavior reflects the complex phase transition dynamics and
complicated contributions from the spin-phonon-electron system.

For a general input at this point, let us recall a  few specific heat data, like
that on $CaMnO_3$ characterized by a $T_N$ = 131 K  \cite{Cornelius}, or that on
\linebreak $La_{0.875}Ca_{0.125}MnO_3$ which presents $three$ well spread anomalies at 315,
146 and 80 K, and a broad (hysteretic) feature at 35K  \cite{Castro}.

Finally, it is well known that the specific heat has simple behaviors at low T,
arising from different mechanisms :(i) the electronic contribution, i.e.
$C_{\epsilon} \simeq T$, (ii) the phonon contribution, as $C_{\omega} \simeq
T^{3}$, and (iii)  the magnon contribution, i.e. $C_{magn} \simeq
T^{3/2}$.\cite{kittel2}  Since the data has been very finely taken we can expect
to treat it as when searching for critical exponents on a log-log plot.  In the
temperature interval so examined three regimes are markedly evident, see Fig. 3.
At $low$ temperature a $3/2$ exponent is observed till $T=94 $K, followed by a
linear temperature regime up to $ T = 156$ K, thereafter followed by a $T^{2/3}$
regime. Okuda et al. \cite{Okuda} have also recently reported that around the M-I
transition  $T^3$ and $T^{1.5}$ components are  observed in the specific heat, in
a $La_{1-x}Ca_{x}MnO_3$ series at low temperature (below 10K).

\section{Discussion}

Since we are mainly dealing with the temperature changes below the transition
temperature(s), it is reasonable to assume that the observed behaviors can be
attributed to a phonon and an electron background, but also to some magnetic
entropy due to the spontaneous magnetization collective fluctuations. We can
write ${\cal F}={\cal F}_M-{\cal F}_e$ for the balance of magnetic ${\cal F}_M$
and electronic ${\cal F}_e$ free energies participating in the  processes under
discussion. The observed magnetization $M$ should result from the minimization of
${\cal F}$. We have discussed elsewhere \cite{us2} that after trivial
rearrangements, the above functional ${\cal F}$ can be cast into a familiar GL
type form describing a second-order phase transition, namely

\begin{equation} {\cal F} [\eta ]=a\eta ^2 +\frac{\beta}{2}\eta ^4-\zeta \eta ^2,
\end{equation} where $\eta$ is the order parameter, in our notations \cite{us2},
the $square$ of the magnetization. As usual, the equilibrium state of such a
system is determined from the minimum energy condition $\partial {\cal
F}/\partial \eta =0$ which yields $\eta _0$ for $T<T_{0}$

\begin{equation} \eta _0^2=\frac{\alpha (T_{0}-T)+\zeta }{\beta }. \end{equation}

This leads to an expression for the total magnetization \begin{equation}
M=M_{av}+M_{fl}^{-}=M_s\left(\eta _{0}^{2}- \frac{\zeta ^2}{3\beta ^2\eta
_{0}^{2}} \right), \end{equation} in terms of the $\zeta$, $\beta $, and $\eta
_0$ parameter of the GL free energy functional. Given the above definitions, the
critical temperatures are related to each other as follows

\begin{equation} T_{MI}=\left(1-\frac{2M_0H_0}{n_eE_k(0,0)-n_iJS}\right)T_C,
\end{equation} with \begin{equation}
T_C=\left(1+\frac{yn_iJS}{n_eE_k(0,0)}\right)T^{*}, \qquad
y=1-\frac{1}{\sqrt{3}}. \end{equation} where  $T^{*}=170K$, the peak of the GMR
which ''normalizes'' the temperature scale in such a  theory.  In the above
equations, $E_k(0,0)=\hbar ^2/2m\xi ^2(0,0)$, where $\xi (T,H)$ is the
field-dependent charge carrier localization length, and $m$ an effective electron
mass. The balance of the exchange $n_iJS$ and localization induced magnetic
$M_sH_0$ energies can also be described by the parameter $z=n_iJS/M_sH_0$; the
mean-field expression for the ''critical field'' is $H_0=3k_BT_C/2S\mu _B$.
Notice that we found $n_e/n_i\simeq2/3$, where $n_i$ and $n_e$ stand for the
number density of localized spins and conduction electrons respectively, from the
competition between the electron-spin exchange $JS$ and the induced magnetic
energy $M_sH_0$ in analyzing the $R(T)$ data \cite{us1,us2}. The analysis of such
data produces $T_{C}=195K$, $JS=40meV$,  instead of 45 meV in \cite{11}.

However we hereby conjecture that the value of the exchange energy $JS$, whence
the transition temperature, as in any mean field theory \cite{27} depends on the
type of bonding, thus on the $Mn-O-Mn$ bond environment, and calculate such an
effect next.

A list of possible $Mn-O-Mn$ bond environments is given in Table I and Table II
with their respective probability of occurrence, assuming statistical
independence of the conditional probabilities, in other words no short range
order. It is easily understood that due to the respective concentrations of $La$,
$Ca$, and $Y$ a few types of $Mn-O-Mn$ bond environments are practically
relevant, each one being characterized by its ''cluster critical temperature''
related to the ''cluster DE-effective exchange integral'', see Eq.(9). Notice
that due to the incommensurability of the ion content with respect to the
available crystallographic site numbers, such ''clusters'' are unavoidable, and
subsist even after repeated annealing, as later checked from the X-Ray spectrum.

The plausible explanation for an intrinsic  material origin of the substructures
observed in $\rho(T)$, rather than an extrinsic one as in \cite{MAJPLPC}, is
provided by the model of Attfield et al. \cite{Attfield}. It is hereby used in
order to evaluate whether there exist such ''environmental effects'' on the local
distortions of the $Mn-O-Mn$ bond, thereby significantly affecting the local
exchange integrals, whence the ''average transition temperature(s)''. Local
distortions leading to specific exchange integrals responsible for magnetic
transitions in (so called clusters) have been parameterised here using two
quantities : (i) the $coherent$ strain parameter $((r_A)^0-<r_A>)^2$, derived
from the expression of the tolerance factor describes the deviation from the mean
cation size; $(r_A)^0$ is the ideal perovskite A cation ionic radius, i.e. 1.30
$\AA$ for $LaMnO_3$ perovskites \cite{whatref} and  $<r_A> $ is the mean radius
of the occurring $A$-site cations; (ii) the statistical variance $\sigma^2$ in
the distributions of ionic radii for each local configuration; the variance
measures the $incoherent$ strain, i.e. the effect of disorder due to the
disparity or mismatch of individual $A$ cation radii. Thereby we indicate that
different samples with the $same$ doping level and tolerance factors can have
quite $different$ transition temperatures.

Among the clusters, we should disregard at once that formed by $4La$'s, indicated
by a ($\dagger$) in Table 1, since this would correspond to a $LaMnO_3$ compound
which is $not$ ferromagnetic \cite{22}. Only the four other main local
configurations,  i.e. according to the above statistical analysis (Table I) they
occur more than 8\% of the time, have thus been taken into account, considering
specific cationic site distributions around each oxygen ion (see Fig. 4): each
one is octahedraly coordinated by two $Mn$ cations and by four $A$ cations in the
plane perpendicular to the $quasi$ linear $Mn-O-Mn$ bridge. In such a first order
(and reasonable) approximation the change in the spin-spin DE-integral, whence
$T_{C}$, is essentially attributed to local strains resulting from oxygen atom
displacements. A plot of $T_{C}$ $vs.$ $\sigma^2$+ $((r_A)^0-<r_A>)^2$ from data
calculated and reported in Table I is shown in Fig.5.  This indicates a marked
inverse square root relationship (Fig.5), (in contrast to Attfield estimated
linear relationship \cite{whatref}, - markedly unphysical) allowing to rank and
confirm the role of each environment on the respective transition temperatures.
E.g. the $3La,1Ca$ cluster corresponds to the  GMR compound
$La_{0.75}Ca_{0.25}MnO_3$ which has a transition temperature near 225K.

Further {\it ab initio} calculations, outside the scope of this paper, would be
useful in order to estimate more exactly, the relative exchange energies for the
clusters drawn in Fig.4.  This would lead to  establish the values the  $Mn-O$
bond lengths and $Mn-O$ bond angles in the $MnO_6$ octahedra as a function of the
environment. It is also known that the electronic band width $W$ depends on the
$Mn-O$ tilt angles \cite{25}. These depend on the $B-O-B$ bond angles and $B-O$
lengths through the overlap integrals between the $3d$ orbitals of the $B$ ion
and the $2p$ orbitals of the $O$ anion. Moreover $W$ controls the critical
temperature through \begin{equation} T_C  \simeq  W  exp(-\gamma E_{JT})/hw
\end{equation} where $w$ is an appropriate optical mode frequency, and $E_{JT}$
the Jahn-Teller (JT) energy (ca. 0.3 eV) \cite{26,Wang}. $W$ should decrease with
increasing temperature \cite{Wang}, leading to electronic localization at high
temperature. This confirms that the discussed magnetic transitions should occur
below the main M-I one indeed.

For completeness, let us point out that as in several reports one might also
attempt to discuss features in terms of JT effects rather than through a DE
formalism. The JT scheme considers the $Mn-O-Mn$ bond geometry $per$ $se$, we
emphasize the role of the bond environment. Both might be highly related of
course.

In the same line of thought, the specific heat regimes below $T_C$ and $T_{MI}$
are thus easily understood from basic solid state physics as recalled here above.
We emphasize the remarkable agreement between the temperature of the anomalies in
$C$ and $d\rho/dT$ thereby illustrating the Fisher-Langer  formula, i.e. $C$
$\sim$ $d\rho/dT$ at second order phase transitions \cite{FisherLanger}.

For completeness again, let us recall that Castro et al. \cite{Castro} attribute
three widely spaced anomaly in the specific heat of $La_{0.875}Ca_{0.125}MnO_3$
(in order of decreasing temperature) (310 K) to  the formation of magnetic
polaron, (146 K) a paramagnetic to ferromagnetic transition, (80 K) a charge
ordering or spin-glass transition, and the 35 K to  a movement of domain walls or
spin reorientation transition.

One point might still be in doubt, i.e. whether the linear $T$ regime is a truly
electronic effect or a $smooth$ crossover. As an argument of a physical effect,
we may argue   that one should distinguish between a $smooth$ and a $sharp $
crossover, the latter  indicating a specific energy, here a temperature or  a
measure of the corresponding effective exchange integral, while the former points
out to disorder or inhomogeneities. In view of the well defined positions of the
94 and 156 K temperatures we might rightly consider $them$ as crossovers. Keeping
coherent with the above analysis we consider therefore that the temperature
interval between these temperatures is the siege of a specific set of phenomena
without any disorder-like effects.  It might be also argued against our
interpretation that the {\it low temperature electronic regime} giving a linear
variation in temperature extends a little bit high  (euphemism !) in temperature.
This would be valid but only if disregarding the marked importance of the
electronic effects in such materials for which the M-I phase transition occurs at
rather $high$ temperature in fact.

Thus we  seem to have accounted for the  observed temperature dependences of the
specific heat in $La_{0.6}Y_{0.1}Ca_{0.3}MnO_3$, in terms of ideas derived from
the DE localization model for GMR materials. All such features are in agreement
with the usual microscopic spin-polaron and electronic localization theories.
Next, concerning the power law exponent less than unity here above observed in
the specific heat temperature behavior the matter is not so trivial. A $C \simeq
T^{2/3}$ law as found at high temperature can be discussed in terms of the
fractal dimension of the conducting network in the so-called insulating phase, as
in Orbach description of random media excitations \cite{Orbach}. It is known that
depending on the excitation spectrum, be it of bosons($b$), i.e. phonons and
magnons, or fermions($f$), i.e. electrons, their density of states depends on the
effective dimensionality $\tilde{\delta}$ of the system, i. e.

\begin{equation} {\cal D}(\omega)  \sim \omega^{(\tilde{\delta}{_{b}-1})}
\end{equation} or

\begin{equation} {\cal D}(\epsilon)   \sim \epsilon^{(\tilde{\delta_{f}}/2) -1}
\end{equation} for bosons and electrons respectively, thus giving respectively a
specific heat behavior like

\begin{equation} C_{\omega}   \sim  T^{\tilde{\delta_{b}}} \end{equation} and

\begin{equation} C_{\epsilon}   \sim  T^{(\tilde{\delta_{f}}-1)/2} \end{equation}
for the phonon and electron contributions in a truly three dimensional system.
Therefore the exponent  $2/3$ can be the signature of a percolation network for
the hopping charge carriers with a (very reasonable) effective dimensionality
$\tilde{\delta}$ = $7/3$ in the less conducting (high temperature) regime. This
is easily contrasted to the linear regime in the (low temperature) conducting
phase, and the appearance of an electronic transition at intermediate
temperatures.

\section{Conclusion}

On one hand, we have observed on a log-log plot three regimes for the specific
heat in a temperature range encompassing the charge carrier localization
temperature $T_{MI}$  and the magnetic transition temperature $T_C$. At not too
low temperature, a collective spin excitation or magnon signature ($C \simeq
T^{3/2}$) is found as for what we could thus call a magnon-polaron regime or
magneto-polaron excitations in such manganites. A simple $T$ law occurs before
the transitions. A $C \simeq T^{2/3} $ law is found at high temperature and is
interpreted in terms of the fractal dimension of the conducting (hopping) network
in the so-called insulating phase.

On the other hand, in the temperature regime near the drastic  magnetic and
charge localization transitions, the report well illustrates the Fisher-Langer
formula, i.e. $C$ $\sim$ $d\rho/dT$ at second order phase transitions
\cite{FisherLanger}. The various $Mn-O-Mn$ bond environments, within the model of
Attfield \cite{whatref}, explain the complex features of the transition region,
though we disagree with Attfield on the  analytical form of the relationship
between the transition temperatures and the strains.

In so doing the temperature dependence of the regimes in the specific heat (and
the electrical resistivity) can be well accounted for as due to electrons {\it
and} so called spin-polarons $and$ magnon-polarons. Thereby we emphasize the role
of $collective$ excitations beyond the mere $localized$ spin (scattering)
background. During the final writing of this paper, we noticed one by  Huhtinen
et al. \cite{Huhtinen} which explains the overall behavior of $\rho(T)$ as a
weighted sum of a variable range hopping mechanism and of itinerant electrons. In
the latter, they use a $T^{4/5}$ electron-magnon interaction term as well in the
$\rho(T)$ fit.

The results are supported by a free energy functional of Ginzburg-Landau (GL)
type describing magnetic phase transitions near such temperatures. We pin point
the possible occurrence of separated and competing phase transitions depending on
the $A$ and $R$ ion distribution near the $Mn-O-Mn$ bond responsible for the GMR
effect. This leads to a formal description of the specific heat as an equilibrium
property. The case of the electrical resistivity temperature derivative can next
be understood within this framework, as due to electron ''critical scattering
processes'' of correlated (localized) spin fluctuations. Notice that we do not
discuss here whether the conduction mechanism differs above and below $T_C$, as
attempted elsewhere in order to explain optical spectra in related materials,
i.e. $La_{0.67-x}Y_{x}Ba_{0.33}MnO_3$ \cite{Bebenin}. Moreover those different
magnetic cluster scattering processes are weakly relevant for the overall
resistance behavior which has a metallic character up to the localization
transition.

It is interesting to re-emphasize that we associate each critical temperature,
with an exchange integral for a specific $finite$ size cluster,  one critical
temperature usually characterizing a $cooperative$ phenomenon. This is an amazing
demonstration of so many observed long range cooperative phenomena due to short
range interactions; see for example a calculation of a diverging susceptibility
for short range spin-spin interactions in \cite{27}. Moreover,  the
cooperativeness is still manifested in the  fact that only the  clusters having a
concentration larger or as large as the order of magnitude of the percolation
concentration  give a remarkable effect. After this paper was reviewed we came
across an interesting study of critical exponents at CMR transitions 
\cite{FurukawaMotome} where the above comment is outlined. We quote a few lines :
''The DE interaction has a distinguishable property compared to ordinary exchange
interactions in spin systems. Effective ferromagnetic interaction is induced by
the kinetics of electrons which favor extended states with ferromagnetic spin
background to gain the kinetic energy. If we integrate out the electron degrees
of freedom to describe the action as a function of spin configurations, it is
necessary to introduce effective long-range two-spin interactions as well as
multiple-spin interactions which depend on sizes and shapes of (the)
ferromagnetic domain structure. The range of the interaction is determined (in
order) to minimize the total free energy for charge and spin degrees of freedom.
As the system approaches the critical point, the magnetic domain structure
fluctuates strongly. Thus, it is highly nontrivial how the DE interaction is
renormalized, whether it is renormalized to a short-range one, or (how) the
long-range and the multiple-spin interactions become relevant to cause the
mean-field-like transition through suppression of fluctuations. The results ...
indicate that the universality class of the ferromagnetic transition in the DE
model is consistent with that of models with short- range interactions. ''

As we mentioned in the introductory part, along with lowering the Curie point,
$Y$ substitution brings about other important effects indeed. Namely, it drives
the magnetic structure closer to a canted AFM phase (which occurs~\cite{7a} at
$T_{AFM}<<T_C$) thus triggering the development of local AFM fluctuations  within
the parent FM matrix. In turn, these fluctuations cause a trapping of spin
polarized carriers in a locally FM environment leading to hopping dominated
transport of charge carriers between thus formed magneto-polarons, for the whole
temperature interval. In the low temperature conducting ferromagnetic phase, the
collective magnon signature has not to be neglected neither on $\rho$ nor $C$.

It is worth noting that the here above experimental and theoretical results
corroborate magneto-transport measurements and subsequent analysis on the
temperature behavior of the MTEP ~\cite{us2} in terms of {\it strong magnetic
fluctuation effects} near the critical temperature(s).

\vskip 1.0cm {\bf Acknowledgments}

Part of this work has been financially supported by the Action de Recherche
Concert\'ees (ARC) 94-99/174. MA and AG thank CGRI for financial support through
the TOURNESOL program. MA and LH thank S. Sergeenkov for discussions and the
University of Li\`ege Research Council for an equipment  grant. We thank  J. C.
Grenet and R. Cauro, Laboratoire de Thermodynamique Exp\'erimentale, Universit\'e
de Nice-Sophia Antipolis, Parc Valrose F-06000 Nice, Cedex 02, France for
providing the samples. Thanks to K. Ivanova for helping us in the submission
process.

\begin{table}[ht] \begin{center} \caption{Statistical table of the percentage of
the most often expected ($> 8.0\%$) [La,Y,Ca]-clusters in the $ABO_3$ structure
for one $La_{0.6}Y_{0.1}Ca_{0.3}MnO_3$;  notations as in Fig.4. A cluster is
considered to be made of four cations placed on the corners of a square itself in
a perpendicular plane to the quasi linear $Mn-O-Mn$ bond. The clusters are ranked
according to their most probable joint frequency of occurrence, assuming
independent probabilities. The displayed frequency does not take into account a
possible short range order for the (*) clusters for which several ion
configurations are possible. The third column gives the sum of the coherent and
incoherent strain parameter $\sigma^2$+ $((r_A)^0-<r_A>)^2$. The corresponding
critical temperature is indicated}

\vskip 1.0cm

\begin{tabular}{|c|||c|c|c|c||} \hline & Cluster type & frequency & total strain
& critical temperature \\ \hline  I &$3 La, 1Ca$ & $25.92$  & 0.009 & 146\\ II
&$2Ca,2La (*)$ & $19.44$ & 0.011 & 130\\ - &$4La$($\dagger$) & $12.96$ & - & - \\
III &$2La,1Ca,1Y$ & $12.96$ & 0.018 &108 \\ IV &$3La,1Y$ & $8.64$ & 0.020 & 94 \\
\hline \end{tabular}

\end{center} \end{table}

\begin{table}[ht] \begin{center} \caption{Statistical table of the percentage of
more rarely expected [La,Y,Ca]-clusters in the $ABO_3$ structure for
$La_{0.6}Y_{0.1}Ca_{0.3}MnO_3$ with notations as in Table I}

\vskip 1.0cm

\begin{tabular}{|c|c||c|c||} \hline Cluster type & frequency & Cluster type &
frequency  \\ \hline $3Ca,1La$ & $6.48$   & $2Ca,1La,1Y$ & $6.48$  \\ $2La,2Y
(*)$ & $2.16$   & $1La,1Ca,2Y$ & $2.16$ \\ $3Ca,1Y$ & $1.08$   & $4Ca
$($\dagger$) & $0.81$ \\ $2Ca,2Y(*) $ & $0.54$   & $1La,3Y $ & $0.24$  \\ $1Ca,3Y
$ & $0.12$   & $4Y$($\dagger$) & $0.01$   \\ \hline \end{tabular} \end{center}
\end{table}

\newpage
\begin{figure}[htb] \epsfxsize=10cm \centerline{\epsffile{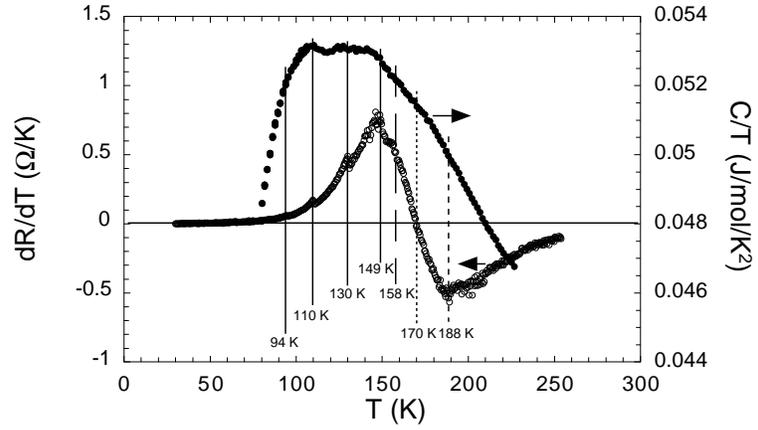} }
\caption{Temperature behavior of the temperature derivative of the electrical
resistivity, i.e. $d\rho (T)/dT$ and the value of $C/T$ where $C$ is the specific
heat of $La_{0.6}Y_{0.1}Ca_{0.3}MnO_3$} \end{figure}

\begin{figure}[htb] \epsfxsize=10cm \centerline{\epsffile{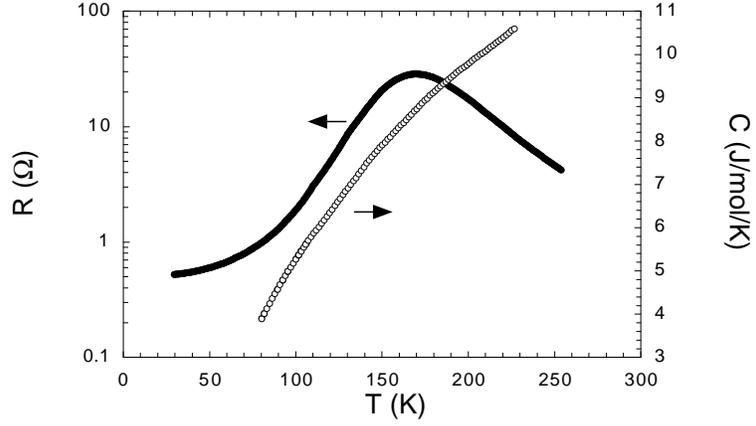} }
\caption{Temperature behavior of the specific heat $C$ and the resistivity $\rho$
at $B=0$ for $La_{0.6}Y_{0.1}Ca_{0.3}MnO_3$ } \end{figure}

\begin{figure}[htb] \epsfxsize=10cm \centerline{\epsffile{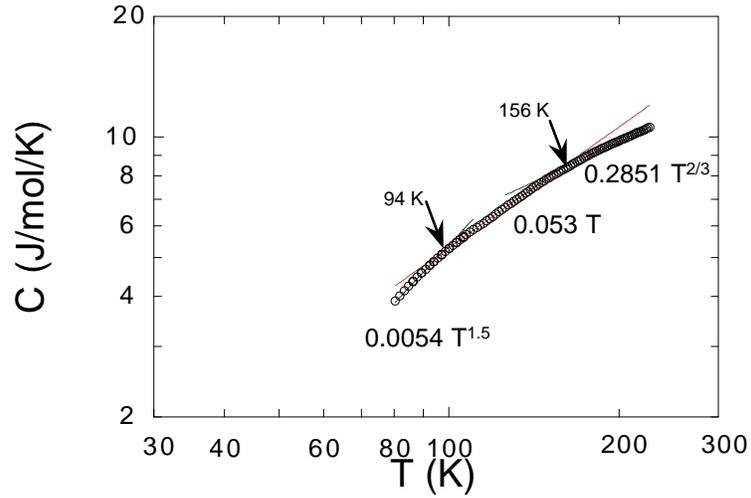} }
\caption{Temperature behavior of the specific heat of
$La_{0.6}Y_{0.1}Ca_{0.3}MnO_3$ on a log-log plot. The best fits to the data
points lead to the indicated slopes } \end{figure}

\newpage
\begin{figure}[htb] 
\caption{Sketch of the four main relevant clusters of $Mn-O-Mn$ bonds in presence
of $La$, $Ca$, and $Y$ at $A$ sites in the $ABO_3$ structure.} \end{figure}

\begin{figure}[htb] 
\caption{Plot  of $T_{C}$  vs. the total coherent and incoherent strain parameter
$\sigma^2$+ $((r_A)^0-<r_A>)^2$ from data collected in \cite{whatref} for the
main relevant clusters of $Mn-O-Mn$ bonds in presence of $La$, $Ca$, and $Y$
distributed on $A$ sites in the $ABO_3$ structure} \end{figure}


\begin{thebibliography}{99}


\bibitem{review}  For a detailed discussion and extensive references, see {\it
Colossal Magnetoresistance, Charge Ordering and Related Properties of Manganese
Oxides}, edited by C. N. R. Rao and B. Raveau (World Scientific, Singapore,
1988); J. M. D. Coey, M. Viret, and S. Von Molnar, Adv. Phys., {\bf 48}, 167
(1999).

\bibitem{7a} A.P. Ramirez, J. Phys. Condens. Matter {\bf 9}, 8171 (1997).

\bibitem{r0}   S. Jin, H.M. O'Bryan, T.H. Tiefel, M. McCormack, and W.W. Rhodes
Appl. Phys. Lett. {\bf 66}, 382 (1995)

\bibitem{1} H.L. Ju, C. Kwon, Qi Li, R.L. Greene, and T. Venkatesan, Appl. Phys.
Lett. {\bf 65}, 2108 (1994).

\bibitem{2} P. Schiffer, A.P. Ramirez, W. Bao, and S.-W. Cheong, Phys. Rev. Lett.
{\bf 75}, 3336 (1995).

\bibitem{3} P.G. Radaelli, D.E. Cox, M. Marezio, S.-W. Cheong, P. Schiffer, and
A.P. Ramirez, Phys. Rev. Lett. {\bf 74}, 4488 (1995).

\bibitem{4} J. Barrat, M.R. Lees, G. Balakrishnan, and D. McPaul, Appl. Phys.
Lett. {\bf 68}, 424 (1996).

\bibitem{5} J. Fontcuberta, M. Martinez, A. Seffar, S. Pinol, J.L. Garcia-Munoz,
and X. Obradors, Phys. Rev. Lett. {\bf 76}, 1122 (1996).

\bibitem{6} J.L. Garcia-Munoz, M. Suaaidi, J. Fontcuberta, and J.
Rodriguez-Carvajal, Phys. Rev. B {\bf 55}, 34 (1997).

\bibitem{7} J. Fontcuberta, V. Laukhin, and X. Obradors, Appl. Phys. Lett. {\bf
72}, 2607 (1998).

\bibitem{8a} J. Fontcuberta, Ll. Balcells, B. Martinez, and X. Obradors, in {\it
Nanocrystaline and Thin Film Magnetic  Oxides} I. Nedkov and M. Ausloos, Eds.,
NATO ASI Series vol. 72 (Kluwer, Dordrecht, 1999) pp. 105-118.

\bibitem{8b} Qi Li and H.S. Wang, in {\it Nanocrystaline and Thin Film Magnetic
Oxides},  I. Nedkov and M. Ausloos, Eds., NATO ASI Series vol. 72 (Kluwer,
Dordrecht, 1999) pp. 133-144.


\bibitem{9} A.J. Millis, P.B. Littlewood, and B.I. Shraiman, Phys. Rev. Lett.
{\bf 74}, 5144 (1995); ibid {\bf 77}, 175 (1996).

\bibitem{10} H. Roder, J. Zang, and A.R. Bishop, Phys. Rev. Lett. {\bf 76}, 1356
(1996).

\bibitem{11} W.E. Pickett and D.J. Singh, Phys. Rev. B {\bf 53}, 1146 (1996).

\bibitem{12} L. Sheng, D.Y. Xing, D.N. Sheng, and C.S. Ting, Phys. Rev. Lett.
{\bf 79}, 1710 (1997).

\bibitem{13} C.M. Varma, Phys. Rev. B {\bf 54}, 7328 (1996).

\bibitem{14} Z.B. Guo, Y.W. Du, J.S. Zhu, H. Huang, W.P. Ding, and D. Feng, Phys.
Rev. Lett. {\bf 78}, 1142 (1997).

\bibitem{us1} S. Sergeenkov, H. Bougrine, M. Ausloos, and A. Gilabert, JETP Lett.
{\bf 70}, 141 (1999)


\bibitem{us2} S. Sergeenkov, H. Bougrine, M. Ausloos, and A. Gilabert, Phys. Rev.
B {\bf 60}, 12 322  (1999)

\bibitem{usPRB} N. Vandewalle, M. Ausloos, and R. Cloots,  Phys. Rev. B {\bf 59},
11909 (1999)

\bibitem{Huhtinen} H. Huhtinen, R. Laiho, K. G. Lisunov, V. N. Stamov and V. S.
Zakhvalinskii, J. Magn. Magn. Mater. {\bf 238}, 160 (2002)

\bibitem{bene} B. Vertruyen, R. Cloots, A. Rulmont, G. Dhalenne, M. Ausloos, and
Ph. Vanderbemden,   J. Appl. Phys. {\bf 90},  5692 (2001)

\bibitem{Castro} M. Castro, R. Burriel, and S.W. Cheong, J. Magn. Magn. Mater.
{\bf 196-197}, 512 (1999)

\bibitem{Okuda} T. Okuda, Y. Tomioka, A. Asamitsu, Y. Tokura, Phys. Rev. B {\bf
61}, 8009 (2000)

\bibitem{Gordon} J. E. Gordon, C. Marcenat, J. P. Franck, I. Isaac, Guanwen
Zhang, R. Lortz, C. Meingast,  F. Bouquet, R. A. Fisher, and N. E. Phillips,
Phys. Rev. B {\bf 65}, 024441 (2001)

\bibitem{Cornelius} A.L. Cornelius, B. Light, J.J. Neumeier,
http://arXiv.org/abs/cond-mat/0108239


\bibitem{Mott} N. F. Mott, Proc. Roy. Soc. A {\bf 153}, 699 (1936); ibid. {\bf
156}, 368 (1936)

\bibitem{VPT}  T. Van Peski-Timbergen and A.J. Dekker, Physica {\bf 29}, 917
(1963).

\bibitem{DGF} P.G. De Gennes and J. Friedel, J. Phys. Chem. Solids {\bf 4}, 71
(1958).

\bibitem{Kasuya} T. Kasuya, Rep. Prog. Phys.  {\bf 16}, 58 (1956); ibid. {\bf
22}, 227 (1959).


\bibitem{MAKD} M. Ausloos and K. Durczewski, Phys. Rev. B  {\bf 22}, 2439 (1980).

\bibitem{MAbk} M. Ausloos,  in {\it Magnetic Phase Transitions}, M.Ausloos and
R.J. Elliott, Eds. (Springer Verlag, Berlin-Heidelberg, 1983) pp. 99-129.

\bibitem{stephaneprb} S. Dorbolo, M. Ausloos and M. Houssa, Phys. Rev. B {\bf
57}, 5401 (1998).

\bibitem{sousa} J.B. Sousa, M.M. Amado, R.P. Pinto, J.M. Moreira, M.E. Braga, M.
Ausloos, J.P. Leburton, J.C. Van Hay, P. Clippe, J.P. Vigneron, and P. Morin, J.
Phys. F {\bf 10}, 933 (1980).

\bibitem{laurent} M. Ausloos and Ch. Laurent, Phys. Rev. B   {\bf 37}, 611
(1988).

\bibitem{FisherLanger}  M.E. Fisher and J.S. Langer, Phys. Rev. Lett. {\bf 20},
665 (1968).

\bibitem{geldart1} D.J.W. Geldart and T.G. Richard, Phys. Rev. B {\bf 12}, 5175
(1975).

\bibitem{geldart2} D.J.W. Geldart, Phys. Rev. B {\bf 15}, 3455 (1977).

\bibitem{kittel2} Ch. Kittel and C. Y. Fong, {\it Quantum Theory of Solids},
Wiley, New York, 1987.

\bibitem{27} H.E. Stanley, {\it Introduction to Phase Transitions and Critical
Phenomena}, Clarendon Press, Oxford, 1968.

\bibitem{10a} P. Wagner, I. Gordon, L. Trappeniers, J. Vanacken, F. Herlach, V.V.
Moshchalkov, and Y. Bruynseraede, Phys. Rev. Lett. {\bf 81}, 3980 (1998).

\bibitem{MAJPLPC} M. Ausloos, J.P. Leburton and P. Clippe,  Solid State Commun.
{\bf 33},  75 (1980)

\bibitem{Attfield} L. M. Rodriguez-Martinez and J. P. Attfield, Phys. Rev. B {\bf
54}, R15622 (1996)

\bibitem{whatref} J. P. Attfield, Int. J. Inorg. Mater. {\bf 3}, 1147 (2001)

\bibitem{22} J.L. Garcia-Munoz, J. Fontcuberta, M. Suaaidi, and X. Obradors, J.
Phys. Condens. Matter {\bf 8}, L787 (1996).

\bibitem{25} M. Melarde, J. Mesot, P. Lacorre, S. Rozenkranz, P. Fischer, and K.
Gobrecht, Phys. Rev. B {\bf 52}, 9248 (1995)

\bibitem{26} G. Zhao, K. Conder, H. Keller, and K.A. Muller, Nature {\bf 381},
676 (1996)

\bibitem{Wang} L. Wang and  X. Zhang, Physica C {\bf 371}, 330 (2002)


\bibitem{Orbach} T. Nakayama, K. Yakubo and R. L. Orbach, Rev. Mod. Phys.  {\bf
66}, 342 (1994)

\bibitem{Bebenin} N.G. Bebenin, N.N. Loshkareva, Yu. P. Sukhorukov, A.P. Nossov,
R.I. Zainullina, V.G. Vassiliev, B.V. Slobodin, K.M. Demchuk, and V.V. Ustinov,
Solid State Commun. {\bf 106}, 357 (1998)


\bibitem{FurukawaMotome} N.Furukawa and Y. Motome, arxiv:cond-mat/0107172; to be
published in  Appl. Phys. A {\bf 75}, xxx (2002)

\end{thebibliography}
\end{document}